\documentclass[journal]{IEEEtran}
\usepackage{multirow}
\usepackage{amsmath,amsfonts,amssymb}
\usepackage{mathrsfs}
\usepackage{mathptmx}
\usepackage{booktabs}
\usepackage{hyperref} 
\usepackage{color}
\usepackage{cite}
\ifCLASSINFOpdf
\usepackage[pdftex]{graphicx}
\graphicspath{{figure/}}
\DeclareGraphicsExtensions{.pdf,.jpeg,.png}
\ifCLASSOPTIONcompsoc
 \usepackage[caption=false,font=normalsize,labelfont=sf,textfont=sf]{subfig}
\else
  \usepackage[caption=false,font=footnotesize]{subfig}
\fi
\begin{document}
\title{
Complex Convolutional Neural Networks\\for Ultrafast Ultrasound Imaging Reconstruction\\from In-Phase/Quadrature Signal
}
\author{Jingfeng~Lu,
        Fabien~Millioz,
        Damien~Garcia,
        S{\'e}bastien~Salles,
        Dong~Ye,
        and~Denis~Friboulet
\thanks{J. Lu is with Metislab, School of Instrumentation Science and Engineering, Harbin Institute of Technology, Harbin, China, and also with the University of Lyon, CREATIS, CNRS UMR 5220, Inserm U1044, INSA-Lyon, University of Lyon 1, Villeurbanne, France (e-mail: jingfeng.lu@hit.edu.cn, jingfeng.lu@creatis.insa-lyon.fr).}
\thanks{F. Millioz, D. Garcia, S. Salles, and D. Friboulet are with the University of Lyon, CREATIS, CNRS UMR 5220, Inserm U1044, INSA-Lyon, University of Lyon 1, Villeurbanne, France.}
\thanks{D. Ye is with School of Instrumentation Science and Engineering, Harbin Institute of Technology, Harbin, China.}
 \thanks{The CID-Net model with trained weights in Open Neural Network Exchange (ONNX) format and a demo script are available at \href{https://github.com/Jingfeng-LU/CID-Net/}{https://github.com/Jingfeng-LU/CID-Net/}}
}
\maketitle
\begin{abstract}
Ultrafast ultrasound imaging remains an active area of interest in the ultrasound community due to its ultra-high frame rates. Recently, a wide variety of studies based on deep learning have sought to improve ultrafast ultrasound imaging. Most of these approaches have been performed on radio frequency (RF) signals. However, in-phase/quadrature (I/Q) digital beamformers are now widely used as low-cost strategies. In this work, we used complex convolutional neural networks for reconstruction of ultrasound images from I/Q signals. We recently described a convolutional neural network architecture called ID-Net, which exploited an inception layer designed for reconstruction of RF diverging-wave ultrasound images. In the present study, we derive the complex equivalent of this network; i.e., the Complex-valued Inception for Diverging-wave Network (CID-Net) that operates on I/Q data. We provide experimental evidence that CID-Net provides the same image quality as that obtained from RF-trained convolutional neural networks; i.e., using only three I/Q images, the CID-Net produces high-quality images that can compete with those obtained by coherently compounding 31 RF images. Moreover, we show that CID-Net outperforms the straightforward architecture that consists of processing the real and imaginary parts of the I/Q signal separately, which thereby indicates the importance of consistently processing the I/Q signals using a network that exploits the complex nature of such signals.

\end{abstract}


\IEEEpeerreviewmaketitle

\section{Introduction}

Ultrafast ultrasound (US) imaging has attracted considerable attention in recent years due to the particularly high temporal resolution that it can provide, which is not readily available with conventional US schemes. In contrast to focused US schemes where successive narrow beams are insonified to scan the entire region of interest, ultrafast US imaging allows a wide field of view to be insonified with a single emission of unfocused transmission beams, such as plane waves \cite{sandrin1999time} and diverging waves (DWs) \cite{hasegawa2011high}. Such imaging modalities allow ultra-high frame rates up to some thousands of frames per second compared with tens of frames per second in line-by-line imaging systems. Due to this high temporal resolution, ultrafast US imaging allows the capture of transient motion and deformation in human tissues, thereby providing new insight into cardiac mechanics, in terms of myocardial mechanical properties \cite{couade2010vivo} and blood flow analysis \cite{bercoff2011ultrafast, faurie2018coupling}.

Ultrafast US imaging is generally based on plane or diverging wavefronts that insonify a broad field of view. The backscattered echoes are then received by the transducer as multi-channel radio frequency (RF) signals. As the incident acoustic energy of the unfocused beams is spread over a broad region, the backscattered echoes received are of low intensity, which results in deterioration of the image quality in terms of contrast and resolution. A standard approach to improve this image quality consists of coherently compounding consecutive backscattered echoes from steered transmit beams at different angles \cite{montaldo2009coherent, tanter2014ultrafast, papadacci2014high, poree2016high}. Compounding-based US imaging can yield images of comparable quality to conventional focusing, at the expense of lower frame rates, larger data volumes, and higher computational costs due to the increased US transmission and processing compared with single plane-wave/diverging-wave imaging. In practice, this inevitably leads to a trade-off between image quality and frame rate. Therefore exploitation of the full potential of ultrafast US imaging in terms of temporal resolution and image quality is a very active area of interest.

Recently, deep learning has promoted growing interest in the US community. In particular, convolutional neural networks (CNNs) have emerged as a promising tool to improve US image formation. A typical class of studies has been devoted to the production of high-quality reconstructions from limited numbers of transmission events associated with high frame rate US imaging techniques, such as plane-wave imaging \cite{gasse2017high, nair2018deep, perdios2018deep, khan2019deep, luijten2019deep, zhou2019ultrafast, perdios2020cnn, zhang2021ultrasound}, DW imaging \cite{lu2019fast, ghani2019high, lu2020reconstruction}, multiline acquisition \cite{senouf2018high, vedula2019learning}, and multiline transmission \cite{vedula2018high}. For example, Gasse et al. \cite{gasse2017high} introduced a CNN to reduce the number of plane wave transmissions for high-quality compounding, by learning a mapping from the images obtained through coherently compounding 31 transmissions. Another type of CNN-based US imaging method aimed at improving image quality from reduced data sampling to increase frame rates\cite{lu2018unsupervised, khan2019deep, yoon2019efficient, khan2020adaptive, huijben2020learning}. For example, Yoon et al. \cite{yoon2019efficient} proposed a CNN to interpolate missing RF channel data from subsampled data, to restore image quality. In addition, the studies of \cite{mishra2018ultrasound, hyun2019beamforming, khan2021variational} used CNNs as denoising filters for speckle reduction in US images.

As US image formation involves a sophisticated signal-processing pipeline, different signal formats along the imaging chain can be used in deep learning, such as RF signals, analytic signals, and demodulated in-phase/quadrature (I/Q) signals. RF data are real-valued amplitude-modulated carrier signals that are obtained from backscattered echoes sensed by US transducers. The US data can be represented as analytic signals, which are obtained by discarding the negative frequency components of the RF signal. Analytic signals are complex valued, where the real components are the original RF signals, and the imaginary components are calculated through the Hilbert transform of the RF signals. Demodulated I/Q data correspond to the complex envelope of the signals and are complex-valued baseband signals that are produced by removing the carrier frequency of the RF signals. Such narrowband setting allows a considerable downsampling factor without aliasing. Hence, in a narrowband setting, it can be advantageous to process and beamform the signals in the baseband using digital I/Q beamformers, as this greatly reduces the data size and power requirement, and therefore the cost at the front-end. 

Among the CNN-based studies in the literature, those of \cite{gasse2017high, nair2018deep, perdios2018deep, luijten2019deep, zhou2019ultrafast, khan2019deep, khan2020adaptive, zhang2021ultrasound, lu2019fast, ghani2019high, lu2020reconstruction, yoon2019efficient, mishra2018ultrasound, khan2021variational} processed RF data using real-valued CNNs. The studies of \cite{khan2019deep, khan2020adaptive,  perdios2020cnn} incorporated analytic I/Q data into the CNN-based imaging pipeline. For example, Khan et al. \cite{khan2019deep, khan2020adaptive} introduced adaptive beamforming models to generate analytic I/Q signals from full or sub-sampled RF signals using a CNN with two output channels, and thus they did not take advantage of the demodulated I/Q signals as the source and output of the network. Only a few studies \cite{hyun2019beamforming, senouf2018high, vedula2019learning, vedula2018high, huijben2020learning} have used demodulated I/Q as the source of the CNN-based imaging. Hyun et al. \cite{hyun2019beamforming} described CNN-based despeckling using demodulated I/Q data to obtain image quality that can compete with the state-of-the-art speckle-reduction algorithms. The CNN was implemented with two input channels that corresponded to real and imaginary parts of the I/Q images. Senouf et al. \cite{senouf2018high} and Vedula et al. \cite{vedula2019learning, vedula2018high}  proposed to improve multiline acquisition and multiline transmission quality by reconstructing images with single-line acquisition quality from demodulated I/Q data. Two real-valued CNNs were trained separately \cite{senouf2018high, vedula2018high} and jointly \cite{vedula2019learning} for the real and imaginary components of the I/Q images.

Although the complex-valued signals can be identified as an ordered pair of real-valued signals in a two-branch network structure, where each branch contains real and imaginary components, such representation does not take into account the nature of the complex calculations. As shown by \cite{hirose2012generalization}, a complex-valued model provides a more constrained system than a model based on real-valued parameters.
A recent study \cite{trabelsi2018deep} demonstrated the superior representational capacity of complex-valued neural networks in acoustic applications such as speech spectrum prediction and music transcription. 


Complex-valued deep learning for US imaging was addressed only very recently by \cite{lu2020complex}, as an introductory version to the present paper. However, the preliminary study did not consider the benefit for the narrowband I/Q signals associated with the much lower sampling rate.
Therefore, inspired by the study of Trabelsi et al \cite{trabelsi2018deep}, here we extend CNN-based US imaging to the complex domain by using a complex CNN (CCNN), thereby allowing direct processing of the demodulated I/Q data, and thus reducing the data size and computational cost. We address, in particular, the following items:

1) We present a CCNN adapted to DW image reconstruction from demodulated I/Q data, called CID-Net (Complex-valued Inception for Diverging-wave Network). CID-Net relies on the complex-valued building components introduced by \cite{trabelsi2018deep}, which allow incorporation of complex numbers into the general frameworks for training deep neural networks. The specific architecture of CID-Net was derived from the inception for DW Network (ID-Net) \cite{lu2020reconstruction}, which has been shown to reconstruct high quality US images using RF data from DW acquisitions. 

2) We devise a specific activation function, as the amplitude \textit{maxout} unit (AMU), for complex-valued activation. As demonstrated by \cite{lu2020reconstruction}, the \textit{maxout} unit (MU) \cite{goodfellow2013maxout} used in conjunction with inception \cite{szegedy2015going} modules contributed to the spatially varying property of ID-Net, which allowed us to deal with the sectorial geometry induced by the DW acquisition. Thus, AMUs are proposed to preserve this property in the complex domain. Also, as will be shown in the following, AMUs allow the maintenance of (i) the complex differentiability of the complex convolutions, and (ii) both the phase and magnitude in the whole complex plane, which is desirable in US imaging reconstruction tasks. 

3) We evaluate the complex network performance to be certain that the data reduction allowed by the CCNN processing of demodulated I/Q data does not impair the reconstructed image quality, as compared to general RF-based CNN reconstruction. We experimentally demonstrate that CID-Net yields the same image quality as that obtained from RF-based CNN, and outperforms the approach that processes the real and imaginary parts of the I/Q signal separately.

The remainder of this paper is structured as follows. In section II, we present the CCNN building blocks and CID-Net for high-quality DW image reconstruction from demodulated I/Q data. The experimental set-up for the data acquisition and network training are presented in Section III. The results are given in Section IV, which are then discussed in Section V. Finally, our concluding remarks are given in Section VI.

\section{Methods}

Let $X\in \mathbb{C}^{m\times w\times h}$ be a complex-valued tensor that represents a limited number ($m$) of beamformed I/Q images from successive DW acquisitions, each of which yields $w$ I/Q signals of length $h$. The DW image reconstruction is modeled as an image input-output problem, where the objective is to estimate a high-quality I/Q image $\hat{Y} \in \mathbb{C}^{w\times h}$ using the low-quality $X$. We propose to use CID-Net with trainable complex-valued parameters $\Theta$ to seek the optimal reconstruction operator $f(\Theta): \mathbb{C}^{m\times w\times h} \mapsto \mathbb{C}^{w\times h}$, with respect to a high-quality target I/Q image $Y\in \mathbb{C}^{w\times h}$ obtained by coherently compounding $\textit{n}$ ($n \gg m$) DW acquisitions. 

\subsection{CCNN Building Blocks}
\subsubsection{Complex convolution}
\begin{figure}[!t]
\centering
\includegraphics[width=3.4in]{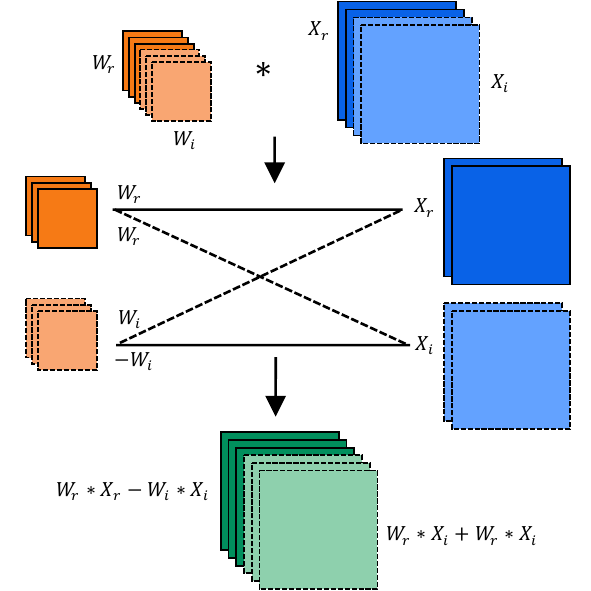}
\caption{Block diagram of complex convolutions. Orange, blue, and green blocks indicate the convolution kernels $W$, input data $X$, and output data $Z$, respectively. Both the data and kernels are formed as the concatenation of two real-valued tensors, each of which represents the real (i.e., solid blocks) and imaginary (i.e., dotted blocks) parts.}
\label{complex conv}
\end{figure}

We followed the representation of complex convolution described by \cite{trabelsi2018deep}, where they used real-valued entities to represent the real and imaginary components of complex numbers, and performed complex convolution using real-valued arithmetic. As shown in Figure \ref{complex conv}, given a complex data tensor $X = X_r + jX_i$ (where $j = \sqrt{-1}$ is the imaginary unit, $X_r = {\rm Re}(X)$ and $X_i = {\rm Im}(X)$ are the real and imaginary components of $X$) and, likewise, complex convolution weights $W = W_r + jW_i$, convolution of $W$ with $X$ yields
\begin{equation}
\label{conv1}
Z = W \ast X = (W_r + jW_i)\ast(X_r + jX_i),
\end{equation}
which reduces to the following when considering the distributive property of convolution, 
\begin{equation}
\label{conv2}
\begin{aligned}
Z = (W_r \ast X_r - W_i \ast X_i) + j(W_r \ast X_i + & W_i \ast X_r).
\end{aligned}
\end{equation}
This representation can be reformulated in algebraic notation as
\begin{equation}
\label{conv3}
\begin{bmatrix}
   {\rm Re}(W \ast X) \\
   {\rm Im}(W \ast X)
\end{bmatrix} = 
\begin{bmatrix}
   W_r & -W_i \\
   W_i & W_r
\end{bmatrix} \ast 
\begin{bmatrix}
   X_r \\
   X_i
\end{bmatrix}.
\end{equation}
Thus the mathematical relations between the real and imaginary components of the data and the convolution weights were fully reflected in this representation, in contrast to a two-branch architecture that would operate separately on the real and imaginary components.

\subsubsection{Activation function}
 
\begin{figure}[!t]
\centering
\includegraphics[width=3.45in]{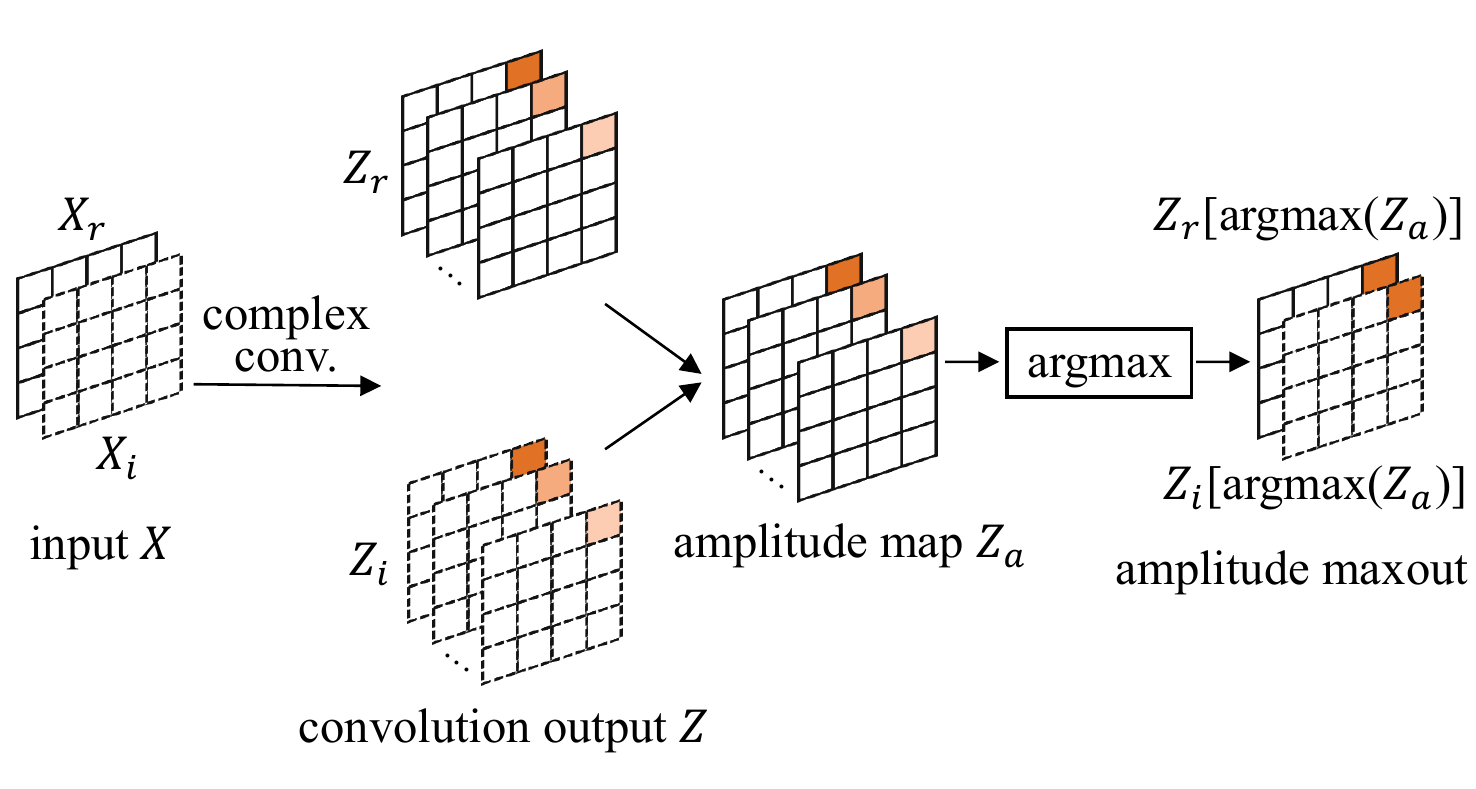}
\caption{Block diagram of the amplitude \textit{maxout} unit (AMU). The solid and dotted blocks indicate the real and imaginary parts of the complex data. An AMU simultaneously activates both the real and imaginary elements corresponding to the element-wise maximum values across the amplitude maps.}
\label{a_maxout}
\end{figure}

 The generalization of the most common activation function, the rectified linear unit (ReLU), to the complex domain is far from being straightforward, as was shown by \cite{trabelsi2018deep}, where three types of ReLU-based complex activations were investigated. As the goal of the present work was to build the complex equivalent of ID-Net with MUs \cite{goodfellow2013maxout} as the activation function, we focused on the design of a complex version of MU activation. In real-valued two-dimensional CNNs, an MU takes the pixel-wise maximum values across several affine feature maps to provide nonlinear transformation. In complex-valued networks, as it is not clear how to determine the maximum value among complex numbers, the maximum operation must be redefined for the complex-valued MU. One simple solution is to apply MU activation to real and imaginary features separately. However, such activation shares the same drawback as two-branch CNNs; i.e., it ignores the interaction between the real and imaginary channels. Thus we devised the AMU for CID-Net. As illustrated in Figure \ref{a_maxout}, given the complex convolution output $Z$, an AMU simultaneously activates the real and imaginary elements that correspond to the element-wise maximum values across the amplitude maps $Z_a = \vert Z \vert$. The AMU can be represented in algebraic notation as
 \begin{equation}
\label{amu}
\begin{bmatrix}
   {\rm Re}({\rm AMU}(Z)) \\
   {\rm Im}({\rm AMU}(Z))
\end{bmatrix} = 
\begin{bmatrix}
   {\rm Re}(Z)[{\rm argmax}(Z_a)] \\
   {\rm Im}(Z)[{\rm argmax}(Z_a)]
\end{bmatrix},
\end{equation}
where ${\rm argmax}(Z_a)$ are the element-wise indices that correspond to the maximum values of $Z_a$ along the channel dimension. 

Two properties of AMU should be noted here: phase encoding and network differentiability. First, compared with the ReLU-based complex activation of \cite{trabelsi2018deep} where the phase was eliminated or modified, AMU preserves both the magnitude and phase of signals throughout the complex plane, as it only activates a convolution to forward-propagate signals in both the real and imaginary channels. Secondly, in the same way, AMU does not change the differentiability of complex convolutions, as it only selects a convolution through which to back-propagate gradients. 

\begin{table*}[!t]
\renewcommand{\arraystretch}{1.4}
\setlength{\tabcolsep}{3.2mm}
\caption{Architectures of ID-Net, CID-Net and 2BID-Net.}
\label{network}
\centering
\begin{tabular}{c c c c c c c c}
\toprule
\multirow{2} * {Layer type} &Feature size& \multirow{2}*{Kernel number} &\multicolumn{2}{c}{Kernel size (height $\times$ width)} & \multicolumn{2}{c}{Activation function}\\ 
~ & (channel $\times$ height $\times$ width) & ~ & ID-Net & CID-Net / 2BID-Net  & ID-Net / 2BID-Net & CID-Net \\
\midrule
Input&$3 \times h \times w$&$-$&$-$&$-$&$-$&$-$\\
Convolution & $64 \times h \times w$ & 256 & $~9 \times 3~$ & $~3 \times 3~$ & 4-piece MU & 4-piece AMU\\
Convolution & $32 \times h \times w$ & 128 & $17 \times 5~$ & $~5 \times 5~$ & 4-piece MU& 4-piece AMU\\
Convolution & $16 \times h \times w$ & 64  & $33 \times 9~$ & $11 \times 9~$ & 4-piece MU& 4-piece AMU\\
\cline{3-7}
\multirow{4} * {Inception} &\multirow{4} * {$8 \times h \times w$} & 8 & $41 \times 11$ & $15 \times 11$ & 4-piece MU& 4-piece AMU\\
~ & ~  & 8 & $49 \times 13$ & $17 \times 13$ & 4-piece MU& 4-piece AMU\\
~ & ~ & 8  & $57 \times 15$ & $19 \times 15$ & 4-piece MU& 4-piece AMU\\
~ & ~ & 8  & $65 \times 17$ & $21 \times 17$ & 4-piece MU& 4-piece AMU\\
\cline{3-7}
Convolution & $1 \times h \times w$ & 4  & $1 \times 1$& $1 \times 1$ & 4-piece MU& 4-piece AMU\\
\bottomrule
\end{tabular}
\end{table*}

\subsubsection{Complex differentiability}

Performing back-propagation in a complex-valued neural network implies differentiable loss functions and activations. One possibility would be to use functions that allow complex derivatives (i.e., holomorphic functions that satisfy the Cauchy-Riemann conditions) \cite{trabelsi2018deep}. This choice is, however, relatively restrictive, and it was shown by \cite{hirose2012generalization} and \cite{trabelsi2018deep}  that such a restriction is not necessary. Indeed, cost functions and activation functions that are differentiable with respect to the real and imaginary parts of each parameter are also compatible with back-propagation.

We used complex mean squared error as the loss function

\begin{equation}
\label{loss}
\begin{aligned}
L(\Theta) =& \frac{1}{n} \sum_{i=1}^n \| \hat {Y_i} - Y_i \|^2.
\end{aligned}
\end{equation}
As the loss function produces real-valued output, it is non-holomorphic and was optimized by adoption of back-propagation with respect to the real and imaginary parts of the convolution weight.

\subsection{Network Architectures}

The CID-Net architecture\footnote{The CID-Net model with trained weights in Open Neural Network Exchange (ONNX) format and a demo script are available at \href{https://github.com/Jingfeng-LU/CID-Net/}{https://github.com/Jingfeng-LU/CID-Net/}} is given in Table \ref{network}, as derived from the ID-Net \cite{lu2020reconstruction} that was recently proposed for RF-based US imaging reconstruction. CID-Net is a two-dimensional convolutional network with five complex convolution layers, which is constructed using the complex convolution blocks and four-piece AMU activation described in Section II-A. 
Specifically, CID-Net is a fully convolutional network with no spatial pooling for signal downsampling. Each layer produces feature maps of the same size using the proper padding of its inputs, which is desirable to preserve the spatial information at the same resolution throughout the network. The four-piece AMU activates pixel-wise the maximum amplitudes across each four affine feature maps produced by the complex convolution. The second to last layer is an inception module that is composed of parallel multi-scale convolutions. Four convolution kernels of different scales are applied to the same feature maps from the previous layer, to obtain feature maps from multi-size receptive fields. The outputs of all of the paths are stacked along the feature dimension as the final output of the inception module. $1 \times 1$ convolution followed by AMU activation is used in the last layer for channel dimension reduction and to generate one-channel reconstructed images. The last layer is equivalent to cross-channel parametric pooling, which can be trained to select element-wise the main elements that contribute to production of the output element. As demonstrated by \cite{lu2020reconstruction}, the inception module used in conjunction with $1\times1$ convolution and \textit{maxout} allows position-dependent features from different receptive fields to be captured, which helps to address the specific geometry of DWs. CID-Net results in multi-size receptive fields (the input patch size for each output pixel) between $31 \times 25$ and $37 \times 31$ (a physical region between 1.2 cm $\times$ 11.7$^{\circ}$ and 1.5 cm$\times$14.5$^{\circ}$, in polar coordinates), and 1.1 million trainable parameters in total.

We also trained an ID-Net model using RF data, as well as a two-branch ID-Net (2BID-Net) model where as in \cite{senouf2018high, vedula2018high}, each branch was trained separately on the real and imaginary parts of the I/Q data. For fair comparisons, the ID-Net, CID-Net, and each branch of the 2BID-Net shared the same feature numbers in each layer. 2BID-Net was used with the same convolutional kernel sizes as CID-Net, while to obtain the same physical receptive field size on RF images, ID-Net used larger sizes in the axial dimension, as indicated in Table \ref{network}. The four-piece MU was used as the activation function, which took pixel-wise max value across each four affine feature maps from the previous convolution layer.

\section{Experiments}

\subsection{Data Acquisition}

A phased array probe (ALT P4-2; bandwidth, 2-4 MHz; center frequency, 3 MHz; pitch, 0.3 mm; kerf, 50 ${\rm\mu}$m)   was interfaced with a research scanner (Vantage 256; Verasonics) to perform steered DW acquisitions. For each acquisition, 31 DWs were emitted with tilt angles between $\pm 30^\circ$ and an incremental step of $2^\circ$. The DWs were transmitted at a pulse repetition frequency of 1500 Hz. The probe was moved manually, on the \textit{in-vitro} and \textit{in-vivo} surfaces, to generate a wide range of significantly different images for the proper training of the network. The received raw RF data were sampled at 12 MHz. These raw RF data were demodulated using a phaser at the center frequency (downmixing). A tenth-order low-pass filter (Butterworth) was then used for bandwidth reduction. The narrowband I/Q signals were sampled at 4 MHz. In both cases (i.e., RF and I/Q data), beamforming was performed according to the delay and sum method described by \cite{perrot2021so}, using a mean sound velocity of 1540 m/s and no apodization. The RF images were beamformed on a $1013 \times 192$ grid, which corresponds to a length of 130 mm (from 10 mm to 140 mm) in the axial direction, and a sectorial region of $90^{\circ}$. In contrast, the I/Q images were beamformed using a $338 \times 192$ grid, which corresponds to the same physical dimension as the RF images.

 From each acquisition, one target image $Y$ was obtained by coherently compounding all $n = 31$ beamformed images, while a small subset of $m$ = 3 beamformed images that corresponded to the steering angles (-20$^{\circ}$, 0$^{\circ}$, 20$^{\circ}$) were used as the network input $X$. A total of 7500 ($X,Y$) samples (i.e., acquisition pairs) were used in the experiment. Specifically, 6000 acquisitions were performed on two \textit{in-vitro} phantoms (Gammex, model 410SCG; CIRS, model 054GS). 1500 acquisitions were performed on \textit{in-vivo} tissues (thigh muscle, finger phalanx, liver regions). These \textit{in-vivo} data were acquired from informed and consenting healthy (non-patient) volunteers.
   
 Besides, we acquired one cardiac sequence in order to show the interest of ultrafast cardiac imaging with DW. The data were acquired with a different center frequency (2.5 MHz) and bandwidth (1-4 MHz), and were sampled in a length of 90 mm (from 10 mm to 100 mm). Hence the cardiac images were not included in the training, and only used for evaluation.

\subsection{Network Training}

In all, 5000 ($X,Y$) samples were randomly selected from the entire dataset as the training set, 1250 ($X,Y$) samples were used as an independent validation set, and the remaining 1250 ($X,Y$) samples were used as the testing set for evaluation. The three models (i.e., 2BID-Net, CID-Net, ID-Net) were trained with the same training implementation, as follows. The network weights were initialized with the Xavier initializer \cite{glorot2010understand}. The loss was minimized using mini-batch gradient descent with the Adam optimizer \cite{kingma2014adam}, and the batch size was set to 16. The initial learning rate was set to $1 \times 10^{-4}$ and an early stopping strategy was used to adjust the learning rate. The learning rate was halved if there had been no decrease in the validation loss for 10 epochs, and the training ended if there were 20 epochs without any validation loss reduction. The training was performed using the PyTorch \cite{paszke2017automatic} library on an NVIDIA Tesla V100 GPU with 32 GB of memory, which resulting in one-day training for CID-Net.

\subsection{Evaluation Metrics}

The imaging quality was quantitatively evaluated using peak signal-to-noise ratio (PSNR), structural similarity index (SSIM), mutual information (MI), contrast ratio (CR), contrast-to-noise ratio (CNR), generalized contrast-to-noise ratio (gCNR) and lateral resolution (LR). PSNR, SSIM, and MI were used as the reference-based indices, to quantify the similarities between the reconstructed images and the reference images (i.e., images obtained through the standard coherent compounding of 31 DWs). The other four metrics (CR, CNR, gCNR, and LR) were reference-free and were used as specialized ultrasound evaluation indices to evaluate the cyst images and the width of the point spread function from point target images. All of the metrics were calculated on the normalized envelope images without log compression.

\subsubsection{Reference-based metrics}

The PSNR is a measure of image quality based on the pixel-based similarities between two images, which is defined as the ratio of the maximum possible value of the image and the distorting noise that deteriorates the quality of its representation. Given the reconstructed image $\hat Y$ and the reference $Y$, the PSNR is calculated as follows.

\begin{equation}
\label{psnr}
\begin{aligned}
{\rm PSNR} &= 20 \log_{10}\frac{{\rm MAX}_Y}{{\rm RMSE}(\hat Y, Y)}\\
&= 20 \log_{10}\frac{{\rm MAX}_Y}{\sqrt{\frac{1}{N}\| \hat Y - Y \|_2^2}},
\end{aligned}
\end{equation}
where ${\rm MAX}_Y$ is the maximum pixel value of the image, RMSE is the root mean square error, $\| \cdot \|_2$ is the ${l}_2$ norm, and $N$ is the number of pixels in the images.

The SSIM measures the structural similarity between the reconstruction $\hat Y$ and the high-quality reference $Y$.

\begin{equation}
\label{ssim}
{\rm SSIM} = \frac{(2\mu _{\hat Y} \mu _{Y} + C_1)(2\sigma _{\hat{Y}{Y}} + C_2)}{(\mu _{\hat Y}^2 + \mu _Y^2 + C_1)(\sigma _{\hat Y}^2 + \sigma _Y^2 + C_2)},
\end{equation}
where $\mu _{\hat Y}$ and $\mu _Y$ ($\sigma^2_{\hat Y}$ and $\sigma^2_Y$) are the means (variances) of $\hat Y$ and $Y$, respectively, $\sigma _{\hat{Y}{Y}}$ is the covariance between $\hat Y$ and $Y$, and $C_1$ and $C_2$ are two constants that stabilize the division with a weak denominator.

The MI reflects the image similarity with the mutual dependence between two images.
\begin{equation}
\label{mi}
{\rm MI} = \sum_{\hat{y},y}p_{\hat{Y}Y}(\hat{y},y){\rm log}\frac{p_{\hat{Y}Y}(\hat{y},y)}{p_{\hat Y}(\hat{y})p_Y(y)},
\end{equation}
where $p_{\hat{Y}Y}(\hat{y},y)$ is the joint distribution of $\hat Y$ and $Y$, and $p_{\hat Y}(\hat y)$ and $p_Y(y)$ are the marginal probability distributions of $\hat Y$ and $Y$, respectively.

For the testing stage, the reference-based metrics were calculated from the full set of testing images.

\begin{figure}[!t]
\centering
\includegraphics[width=2.6in]{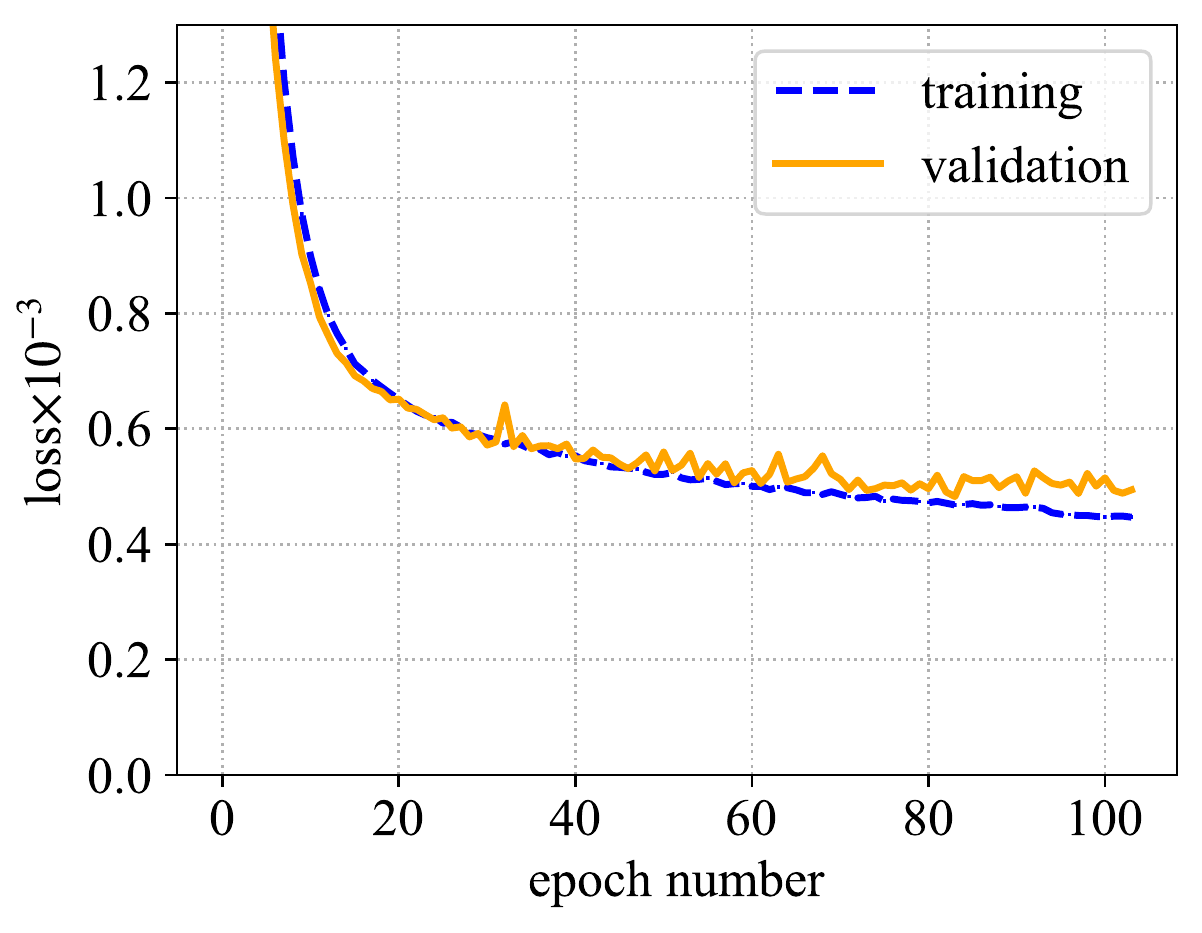}
\caption{Training loss (solid blue curve) and validation loss (dotted orange curve) of CID-Net in the training stage.}
\label{loss_curve}
\end{figure}

\begin{figure*}[!t]
\centering
\includegraphics[width=2.05\columnwidth, angle=0]{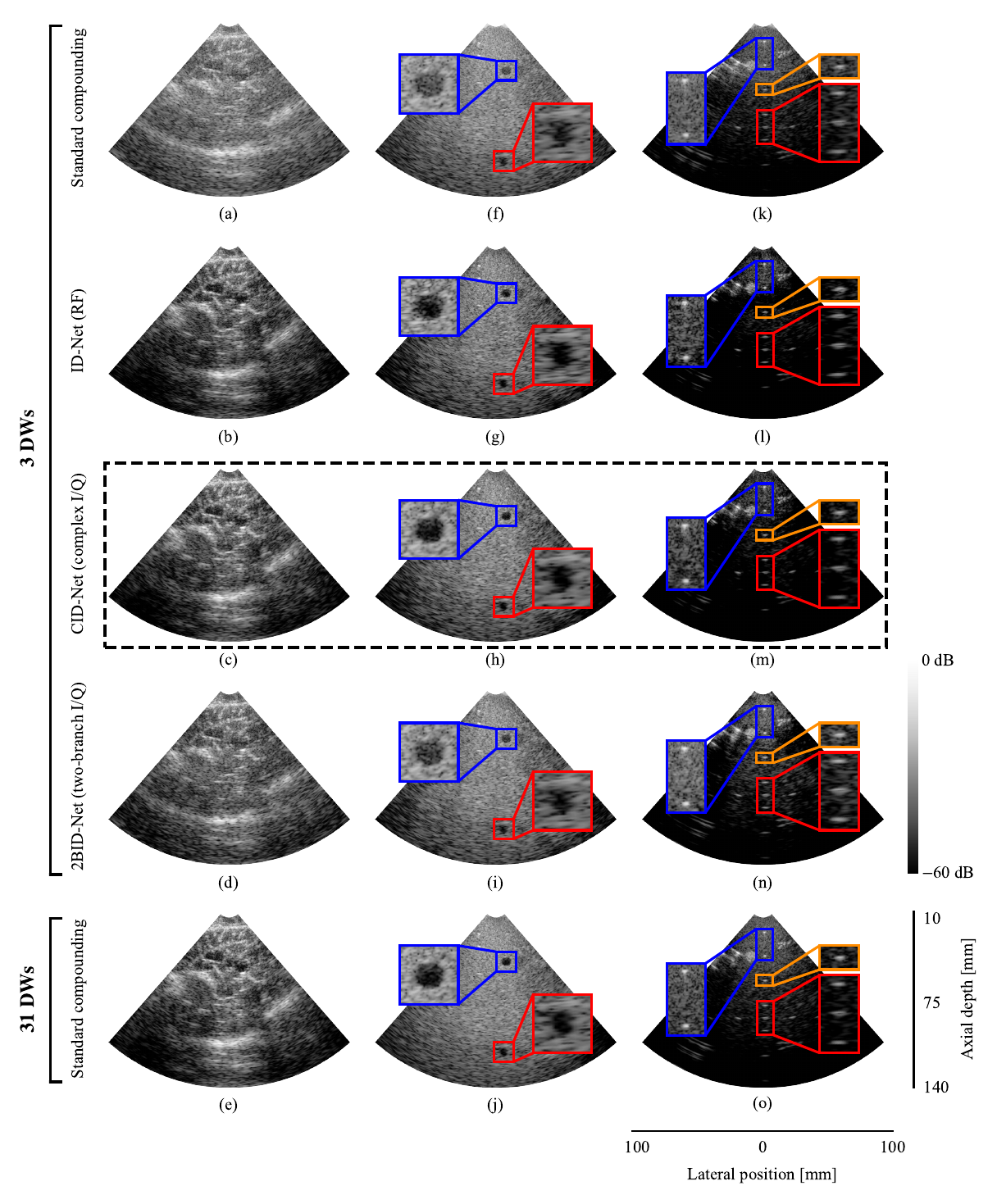}
\caption{B-mode images obtained using different methods. Left to right: \textit{in-vivo} tissues from the thigh muscle; \textit{in-vitro} tissues from the phantom (Gammex, model 410SCG); and \textit{in-vitro} tissues from the phantom (CIRS, model 054GS). Top to bottom: standard coherent compounding of 3 DWs; reconstruction of ID-Net (3 RF images), CID-Net (3 I/Q images), and ID-Net (3 RF images); and standard coherent compounding of 31 DWs (reference).}
\label{bmode}
\end{figure*}

\subsubsection{Reference-free metrics}
 
The CR, CNR, and gCNR were used to measure the contrast between the object of interest and the surrounding background.
 
\begin{equation}
\label{cr_formula}
{\rm CR} = -20 \log_{10}\frac{ \mu _t}{\mu _b},
\end{equation}
\begin{equation}
\label{cnr_formula}
{\rm CNR} = 20 \log_{10}\frac{\vert \mu _t - \mu _b\vert}{\sqrt{\sigma^2_t + \sigma^2_b}},
\end{equation}
where $\mu _t$ and $\mu _b$ ($\sigma^2_t$ and $\sigma^2_b$) are the means (variances) of the intensities within the target region and the background, respectively. 

The gCNR was recently introduced as a generalized variant of CNR for unbiased contrast measures
 It is defined as the non-overlapping proportion between the intensity distribution within two regions.
\begin{equation}
\label{gcnr_formula}
{\rm gCNR} = 1- \sum_k min\{p_t(k), p_b(k)\},
\end{equation}
where $k$ is the pixel intensity, and $p_t(k)$ and $p_b(k)$ are the probability mass functions of the pixel intensity $k$ within the target and background regions.

The LR was used to assess the width of the point spread function from point target images. The full width at half maximum (FWHM) was used in this work to estimate the LR.

In the testing stage, the CR, CNR, and gCNR were measured on two anechoic regions (in the near field at 40-mm depth, and in the far field at 120-mm depth) of an image obtained from a phantom (Gammex, model 410SCG). The LR was measured on 0.1-mm Nylon monoﬁlaments (in the near field at 20-mm and 40-mm depth, in the middle field at 60-mm depth, and in the far field at 80-mm, 90-mm, and 100-mm depths) of an image obtained from a phantom (CIRS, model 054GS).

\begin{table}[!t]
\renewcommand{\arraystretch}{1.4}
\setlength{\tabcolsep}{1.2mm}
\caption{PSNR, SSIM, and MI reached by different methods on the testing data.}
\label{reference_metrics}
\centering
\begin{tabular}{c c c c}
\toprule
Model & PSNR [dB] &  SSIM & MI \\ 
\midrule
Standard compounding (3 DWs)  & $29.68 \pm 1.24$  & $0.80 \pm 0.14$   & $0.58 \pm 0.16$\\
ID-Net (3 RF images)  & $\mathbf{31.57\pm 1.25}$ & $\mathbf{0.92\pm0.06}$  & $\mathbf{0.81 \pm 0.23}$\\
CID-Net (3 I/Q images)  & $31.56\pm 1.39$   & $\mathbf{0.92\pm 0.06}$ & $\mathbf{0.81\pm 0.24}$\\
2BID-Net (3 I/Q images)  & $30.47\pm 1.35$   & $0.86\pm 0.11$   & $0.73 \pm 0.25$  \\
\bottomrule
\end{tabular}
\end{table}

\begin{figure}[!t]
\centering
\includegraphics[width=\columnwidth]{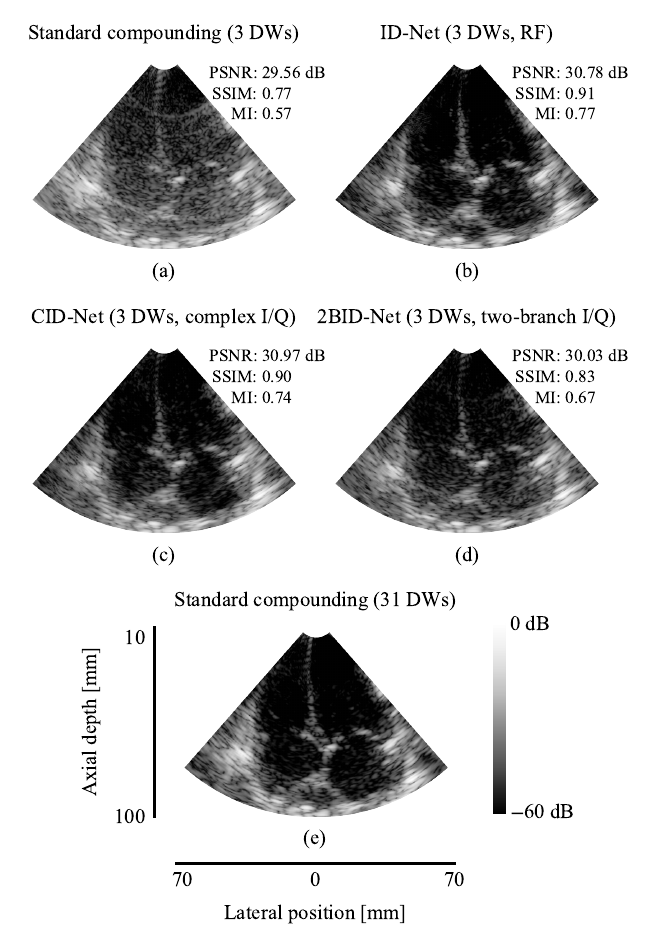}
\caption{Cardiac images obtained from standard coherent compounding of 3 DWs (a); reconstruction of ID-Net (b), CID-Net (c), and 2BID-Net (d); and standard coherent compounding of 31 DWs (e).}
\label{cardiac}
\end{figure}

\begin{table*}[!t]
\renewcommand{\arraystretch}{1.4}
\setlength{\tabcolsep}{2.4mm}
\caption{CR, CNR, gCNR, and LR reached by different methods on the testing data.}
\label{metrics}
\centering
\begin{tabular}{c c c c c c c c c c}
\toprule
\multirow{2} * {Model} & \multicolumn{2}{c}{CR [dB]}  & \multicolumn{2}{c}{CNR [dB]} & \multicolumn{2}{c}{gCNR} & \multicolumn{3}{c}{LR [mm]} \\ 
~ & near field & far field & near field & far field & near field & far field & near field & middle field& far field \\
\midrule
Standard compounding (3 DWs) & 11.61 & 11.82 & 3.42 & 2.96 & 0.747 & 0.703 & 0.85 & 1.52 & 2.26 \\
ID-Net (3 RF images) & \textbf{21.08}  & \textbf{18.90}  & 8.45 & 6.00 & \textbf{0.930} & 0.811 & \textbf{0.99} & 1.81 & \textbf{2.56}\\
CID-Net (3 I/Q images)  & 20.90  & 18.74   & \textbf{8.48} & \textbf{6.12} & 0.927 & \textbf{0.815} & 1.01 & \textbf{1.78} & 2.57\\
2BID-Net (3 I/Q images)  & 19.48 & 16.45 & 7.26 & 4.73 & 0.906 & 0.741 & 1.13 & 1.99 & 2.66\\
 Standard compounding (31 DWs) & 21.53 & 18.84 & 8.62 & 6.30 & 0.934 & 0.833 & 1.04 & 1.89 & 2.61\\
\bottomrule
\end{tabular}
\end{table*}

\begin{figure*}[!t]
\centering
\subfloat[]{\includegraphics[height=0.67\columnwidth]{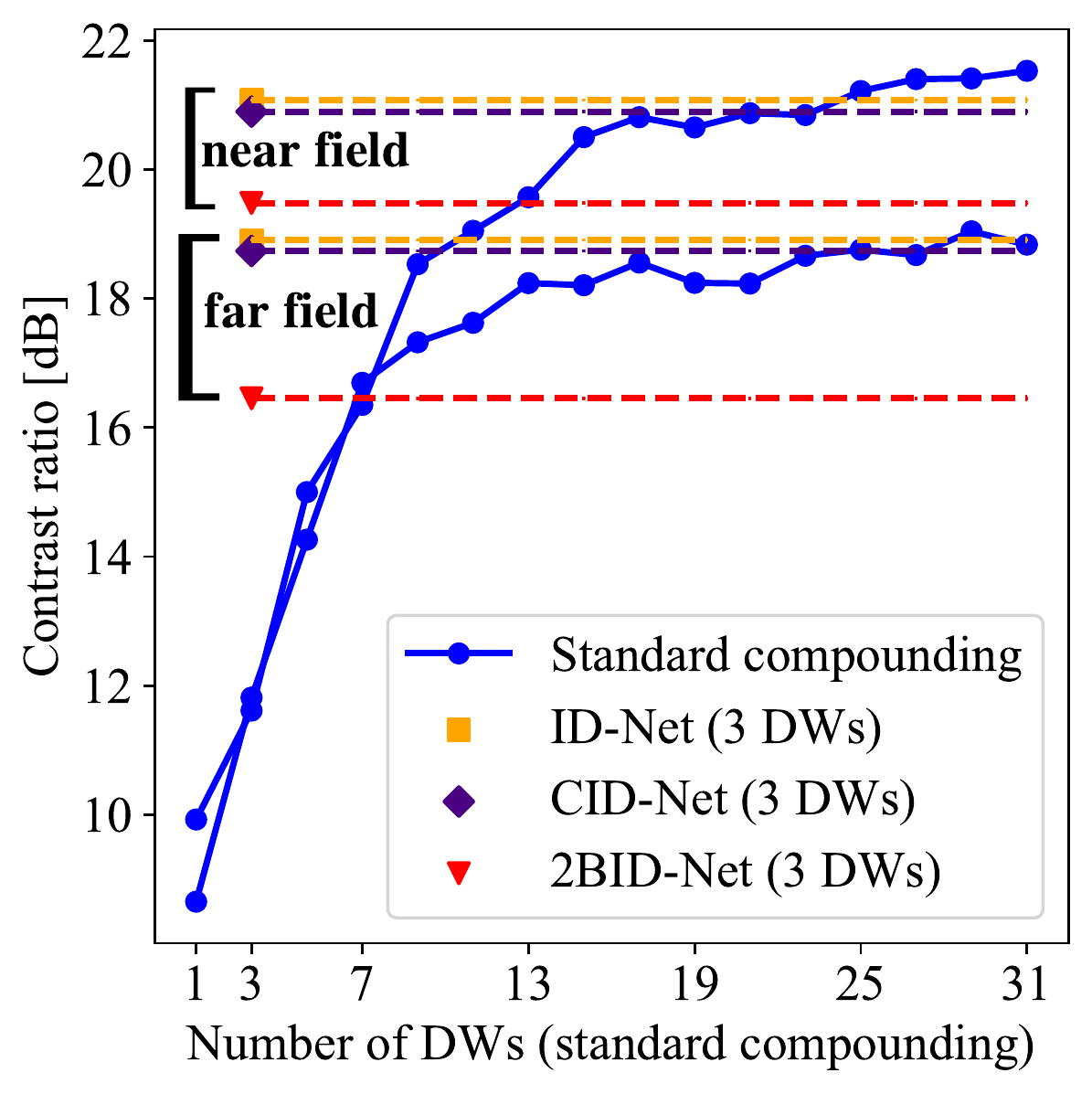}
\label{cr}}
\subfloat[]{\includegraphics[height=0.67\columnwidth]{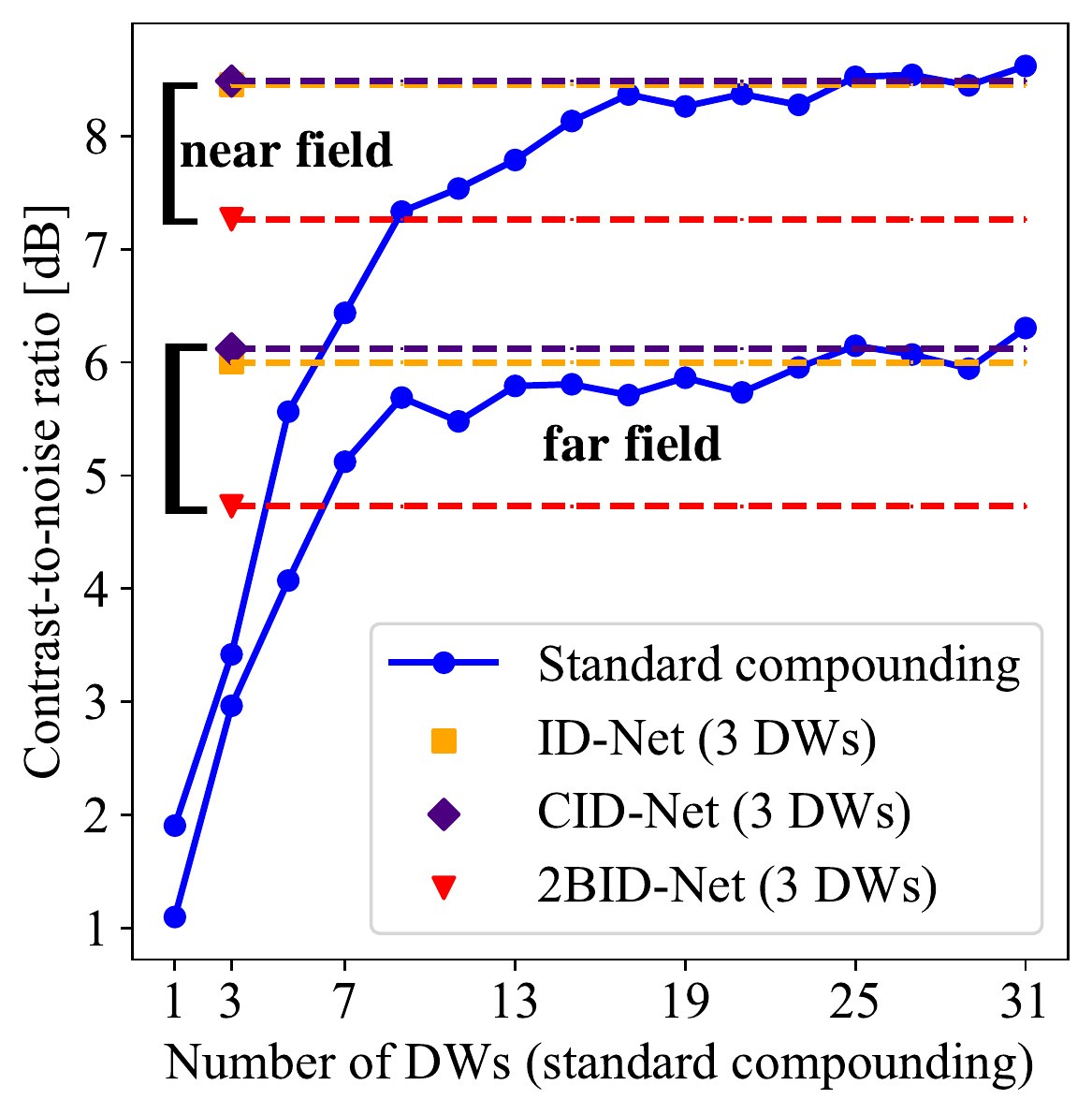}
\label{cnr}}
\subfloat[]{\includegraphics[height=0.67\columnwidth]{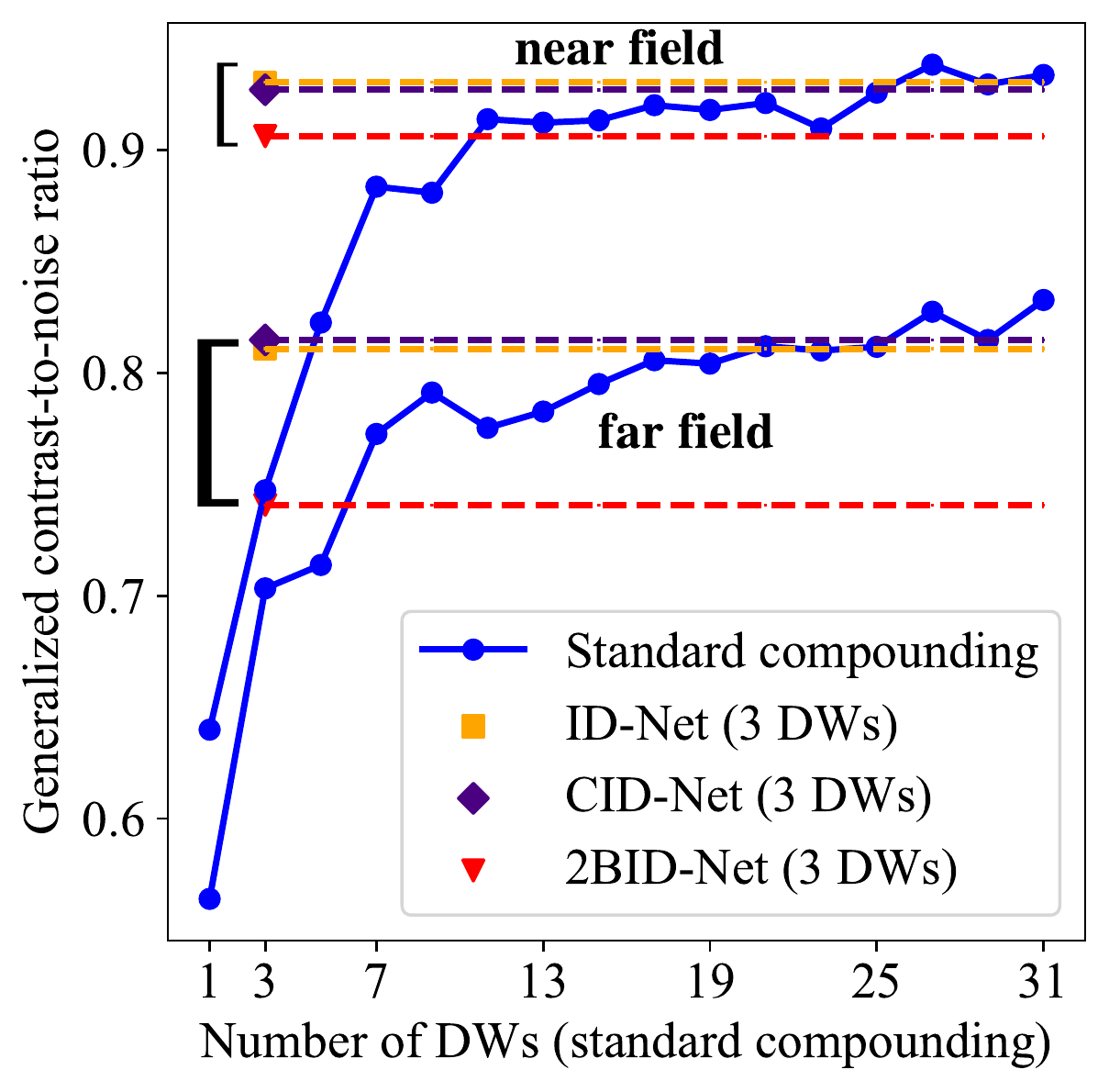}
\label{gcnr}}
\caption{Contrast ratio (a), contrast-to-noise ratio (b), and generalized contrast-to-noise ratio (c) reached by ID-Net (orange squares), CID-Net (purple diamonds), 2BID-Net (red triangles), and standard coherent compounding (blue points and curves). ID-Net, CID-Net, and 2BID-Net used a constant number of three DWs, whereas the results of the standard coherent compounding are given for an increasing number of DWs. The results are given in the near and far fields.}
\label{cr_figures}
\end{figure*}

\section{Results}

\subsection{Learning Convergence}
Figure \ref{loss_curve} shows the evolution of the training loss and validation loss in the training stage. The losses decreased in the early epochs and stabilized to a plateau after around 80 epochs. After 20 more epochs without validation loss reduction, the training consequently ended according to the early stopping strategy described in Section III-B. The training and validation losses showed similarly evolving tendencies, which provide evidence that the training did not suffer from overfitting. Besides, a similar observation held for the training trend of ID-Net and 2BID-Net with the same hyperparameter setting as CID-Net, costing 5.5-day and one-day training time, respectively.

\subsection{Performance of Proposed Method}

\subsubsection{Overall image quality}

Figure \ref{bmode} shows the representative test images from \textit{in-vivo} scans on thigh muscle (first column) and \textit{in-vitro} phantoms (second and third columns), displayed in B-mode with a dynamic range of 60 dB. It appears that the deep-learning-based methods (i.e., ID-Net, second row; CID-Net, third row;  2BID-Net, fourth row) produce better image qualities than those of the standard coherent compounding with the same three DWs. In particular, CID-Net and ID-Net both show significant improvement in contrast and anatomical structures, yielding images that are visually very close to the reference images (with standard coherent compounding of 31 DWs), while 2BID-Net shows a noticeable difference as compared to the reference images. 

To quantitatively assess the performances in terms of visual quality, we report in Table \ref{reference_metrics} the PSNR, SSIM, and MI for each compounding experiment. These reference-based visual metrics were computed from the full testing set, using the standard coherent compounding of 31 DWs as the reference. Consistent with the observations in Figure \ref{bmode}, ID-Net, CID-Net, and 2BID-Net provide higher metric values than those of the standard coherent compounding of the same three DWs. In particular, CID-Net obtains approximately the same values as ID-Net, and higher than those associated with 2BID-Net (gain of 1.09 dB in PSNR, 0.06 in SSIM, and 0.08 in MI). 

Figure \ref{cardiac} shows the reconstructed cardiac images using different methods, and the visual metrics of each reconstruction using Figure \ref{cardiac}e as the reference. A similar visual observation holds for the reconstructed cardiac images as those in Figure \ref{bmode}. In the same way, CID-Net provides close PSNR, SSIM, and MI (30.97 dB, 0.90, and 0.74) as those of ID-Net (30.78 dB, 0.91, and 0.77), and outperforms the standard coherent compounding of 3 DWs (29.56 dB, 0.77, and 0.57) and 2BID-Net (30.03 dB, 0.83, and 0.67). Moreover, the results reached by CID-Net on the cardiac images do not drop much (a decrease of 0.59 dB in PSNR, 0.02 in SSIM, and 0.07 in MI) as compared with those in Table \ref{reference_metrics} obtained with the testing set, which implies a good generalization of the proposed approach under different imaging configurations. 

\begin{figure}[!t]
\centering
\includegraphics[width=2.55in]{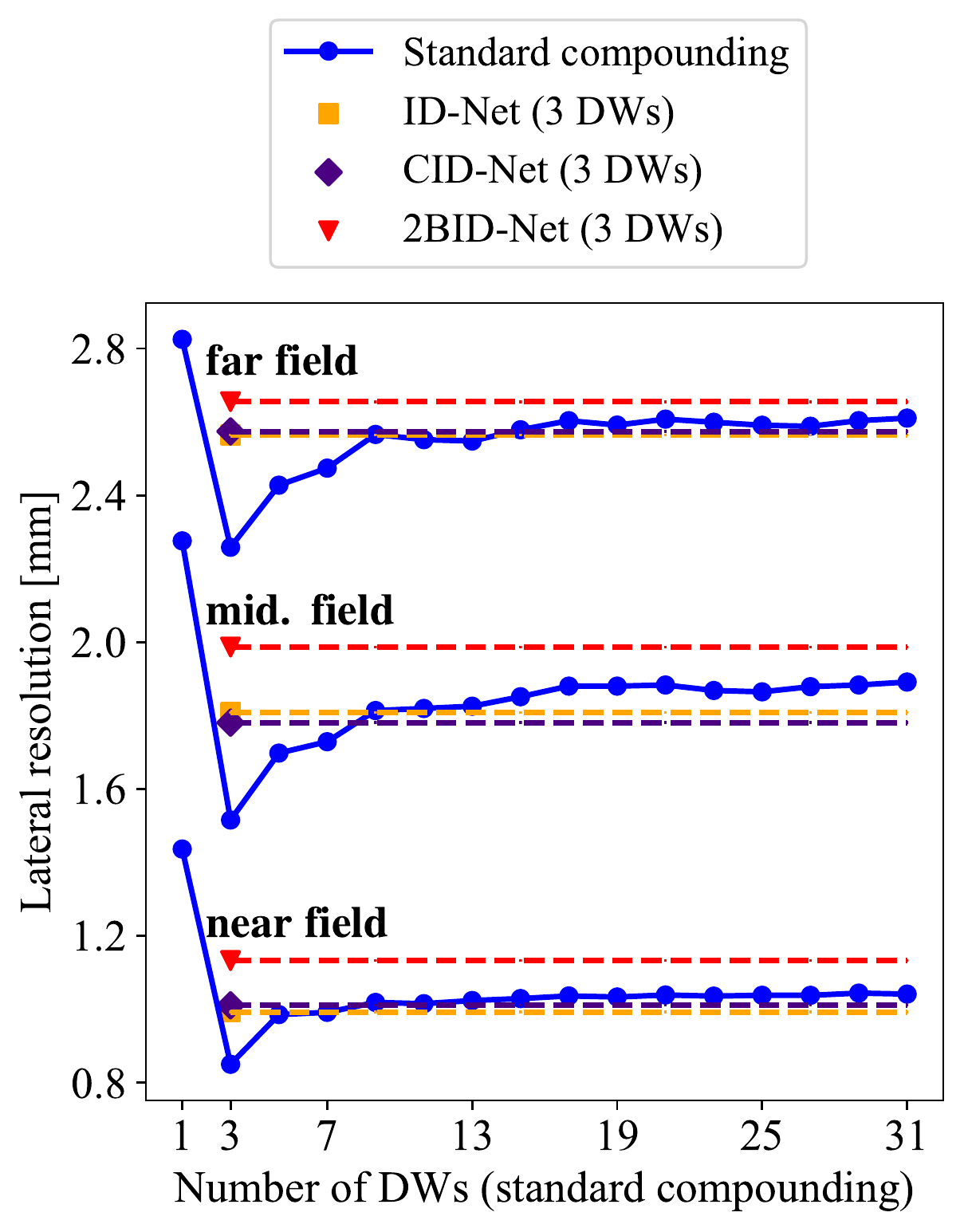}
\caption{Lateral resolution reached by ID-Net (orange squares), CID-Net (purple diamonds), 2BID-Net (red triangles), and standard coherent compounding (blue points and curves). ID-Net, CID-Net, and 2BID-Net used a constant number of three DWs, whereas the results of standard coherent compounding are given for an increasing number of DWs. The results are given in the near, middle, and far fields.}
\label{lrfigures}
\end{figure}

\subsubsection{Image contrast}

As shown from Figure \ref{cr_figures} and Table \ref{metrics}, we investigated the quality of the reconstructed images in terms of image contrast. The contrast was measured for the two anechoic regions in the near field at 40-mm depth and the far field at 120-mm depth, with the images shown in Figure \ref{bmode} (second column). Figure \ref{bmode}i shows loss of contrast in the image produced by 2BID-Net compared with the reference image (Fig. \ref{bmode}j). CID-Net (Fig. \ref{bmode}h) and ID-Net (Fig. \ref{bmode}g) show equivalent visual contrast compared to the reference image (Fig. \ref{bmode}j). Figure \ref{cr_figures} shows the CR, CNR, and gCNR for each compounding experiment. The deep-learning-based methods (ID-Net, CID-Net, 2BID-Net) used a constant number of three DWs, whereas the results of the standard coherent compounding are given for an increasing number of DWs. It can be seen that CID-Net and ID-Net reach approximately the same CR, CNR, and gCNR, which are equivalent to those of the coherent compounding of 25 DWs, in both the near and far fields. 2BID-Net reaches values comparable to those associated with the coherent compounding of 7 to 13 DWs.

Table \ref{metrics} shows the CR, CNR, and gCNR values reached by the deep-learning-based methods, as well as for the standard coherent compounding of 3 DWs and 31 DWs. Compared with 2BID-Net, CID-Net provides higher CR values (gain of 1.42 dB and 2.29 dB in the near and far fields, respectively), and the same holds for the CNR (gain of 1.22 dB and 1.39 dB in the near and far fields, respectively) and gCNR (gain of 0.021 and 0.074 in the near and far fields, respectively). The results obtained with CID-Net and ID-Net are relatively close, and slightly lower than those for the reference (standard coherent compounding of 31 DWs).

\begin{table*}[!t]
\renewcommand{\arraystretch}{1.4}
\setlength{\tabcolsep}{3mm}
\caption{Axial sample number, parameter number, floating point operations (FLOPs), training time, and inference time associated with 2BID-Net, CID-Net, and ID-Net.}
\label{speed}
\centering
\begin{tabular}{c c c c c c c}
\toprule
\multirow{2} * {Model} & \multirow{2}*{Axial sample number} & \multirow{2}*{Real parameter number} & \multirow{2}*{FLOPs} &   \multirow{2}*{Training time} & Inference time & Inference time \\
~ & ~ & ~ &~ & ~ &  (CPU) & (GPU)\\ 
\midrule
ID-Net & 1013 & 1.7 million & 23.8 billion & 5.5 day & 2.85 s  & 0.76 ms\\
CID-Net & \textbf{338} & \textbf{1.1 million} & 7.0 billion & \textbf{1 day}  & 1.56 s &  0.70 ms\\
2BID-Net & \textbf{338} & \textbf{1.1 million} & \textbf{3.5 billion} & \textbf{1 day} & \textbf{0.80 s}  & \textbf{0.65 ms}\\
\bottomrule
\end{tabular}
\end{table*}

\subsubsection{Lateral resolution}

Figure \ref{lrfigures} and Table \ref{metrics} show the quality of the reconstructed images in terms of LR. The LR was measured as FWHM of the point spread function associated with the scatterers (in the near field at 20-mm and 40-mm depth, the middle field at 60-mm depth, and the far field at 80-mm, 90-mm and 100-mm depth) of the images shown in Figure \ref{bmode} (third column). Figure \ref{lrfigures} shows the LR for each compounding experiment. The deep-learning-based methods (ID-Net, CID-Net, 2BID-Net) used a constant number of three DWs, whereas the results of the standard coherent compounding are given for an increasing number of DWs. Table \ref{metrics} gives the specific LR values reached by the deep-learning-based methods, as well as for the standard coherent compounding of 3 DWs and 31 DWs. From Figure \ref{lrfigures}, the LR of the standard coherent compounding shows a similar evolving tendency as the experimental observation from Zhang et al. \cite{zhang2016extension}: the LR value quickly decreases with two or three DWs, and then reaches a plateau with more DWs. From Table \ref{metrics}, the LR associated with CID-Net is better than that provided by 2BID-Net (decrease of 0.12 mm, 0.21 mm, and 0.09 mm in the near, middle, and far fields, respectively). Also, CID-Net and ID-Net provide approximately the same results as for LR, which are equivalent to those of the coherent compounding of 9 DWs, and lower than those of the reference.

\subsection{Computational Complexity and Speed}

Table \ref{speed} gives the data for each model, in terms of input data size, parameter number, floating point operations (FLOPs), and compounding time on the CPU and GPU (i.e., inference time executed with PyTorch on the platform described in Section III-B). The data in Table \ref{speed} indicate that CID-Net and 2BID-Net processed the I/Q signals with 338 axial samples each line, as fewer than those for ID-Net with the RF signals (1013 axial samples each line). Due to the reduced convolutional kernel size in each layer, CID-Net and 2BID-Net had 1.1 million parameters, which were fewer than those of ID-Net (1.7 million). With their smaller parameter numbers and axial sample numbers, CID-Net and 2BID-Net required fewer FLOPs (7.0 billion and 3.5 billion, respectively), and costed less training time (1 day) and CPU inference time (1.56 s and 0.80 s, respectively) compared with ID-Net (23.8 billion FLOPs, 5.5 days for the training time, and 2.85 s for CPU inference time). It can nevertheless be noted that the IQ-based models (CID-Net and 2BID-Net) provided little improvement in the inference time on the GPU compared with the RF-based ID-Net. This is linked to the parallelism of the convolution operations on the GPU. All models required less than 0.8 ms for inference on the GPU, which allows real-time imaging.

\section{Discussion}
 
I/Q signals are ubiquitous in US imaging processing tasks. Modern US systems typically sample the demodulated I/Q signals instead of the amplitude-modulated RF signals, so as to reduce the data size and computational costs. With the recent progress in deep-learning methods for US imaging, we have considered here the deep-learning methods for I/Q-based US reconstruction. 

To this end, we propose to extend CNN-based US imaging to the complex domain, and here describe a complex-valued CNN (i.e., CCNN) for high-quality DW imaging reconstruction from I/Q signals. The image reconstruction was modeled as an I/Q signal input-output problem, which was solved by the proposed CID-Net described in Section II. The CID-Net was trained to learn an optimal I/Q signal compounding operator of three consecutive DW acquisitions, with respect to high-quality targets obtained from standard coherent compounding of 31 DWs. In Section IV, we investigated the performance of CID-Net in terms of learning convergence, image quality, and computational complexity. Section IV-A shows that the CID-Net successfully converges in the training, with a training time of 1 day. The reconstructed images shown in Figures \ref{bmode} and \ref{cardiac} show that the CID-Net provides high-quality reconstructions, which are visually close to the ground truth. This observation is further demonstrated by the comprehensive quantitative results given in Tables \ref{reference_metrics} and \ref{metrics}. Furthermore, Figures \ref{cr_figures} and \ref{lrfigures} indicate that CID-Net provides equivalent image quality in terms of contrast and resolution, reaching the same CR, CNR, and gCNR as the standard coherent compounding of 25 DWs, and an FWHM close to the reference. From Table \ref{speed}, CID-Net requires 0.7 ms for inference, which allows real-time imaging on the GPU platform.
These results demonstrate the feasibility of applying CCNNs for real-time I/Q-based ultrafast US imaging. Several points of discussion on the results presented here are important to note.

First, we compared the proposed CID-Net with the RF-based ID-Net described by \cite{lu2020reconstruction}. CID-Net is indeed the complex-valued version of ID-Net, as it was constructed by replacing the conventional convolution and MU activation with the complex-valued convolution and AMU, as described in Sections II-A and II-B. The experimental results in Figures \ref{bmode} and \ref{cardiac} demonstrate that CID-Net and ID-Net provide equal image quality, with approximately the same values for all of the evaluation metrics in Tables \ref{reference_metrics} and \ref{metrics}. With equal performance to the RF-based counterpart, ID-Net, the interest in the processing of demodulated I/Q data using complex-valued networks is specifically reflected in two aspects, as follows. 

\subsubsection{Data volume}

Processing I/Q data allows considerable axial downsampling without aliasing, as quadrature demodulation removes the carrier frequency. As shown in Table \ref{speed}, the sample number (338 axial samples) of I/Q data used in the experiment is lower by a factor of three relative to the RF data (1013 axial samples). 

\subsubsection{Model complexity}

The reduced I/Q image dimension requires smaller convolutional masks in each layer to cover the same physical receptive field size compared with real-valued convolutions that operate on RF images, as shown in Table \ref{network}. This leads to decreased neuron numbers in each layer, and consequently lower parameter numbers in total, as shown in Table \ref{speed}. 

The decreased data size and parameter numbers contribute to lower computational costs for CID-Net for both training and inference (1 day for training; 1.56 s and 0.70 ms for inference time on the CPU and GPU, respectively), as compared with ID-Net (5.5 days for training; 2.85 s and 0.76 ms for inference time on the CPU and GPU, respectively). It should nevertheless be noted from Table \ref{speed} that CID-Net provides marginal improvement to the GPU inference speed over ID-Net (decrease of 0.06 ms in inference time), which is linked to the parallelism of the convolution operations on the GPU. To check this aspect, we performed inference using CPU computation, and this indeed provides a faster CPU inference time for CID-Net, which reflects the computational cost reduction using I/Q-based imaging.

Secondly, we compared the proposed CID-Net with a two-branch ID-Net model (i.e., 2BID-Net) where each branch was trained separately on the real and imaginary parts of the I/Q data. As we observed in Section IV, CID-Net provides better image quality compared to 2BID-Net, although both models were trained with the same I/Q data and had the same model size (with the same convolution kernel sizes and kernel numbers in each layer, and consequently the same parameter number in total). This was achieved by the complex convolution described in Section II-A. As described in Equation (\ref{conv2}), the complex convolution requires four convolution computation, as $W_r \ast X_r$, $W_i \ast X_i$, $W_r \ast X_i$, and $W_i \ast X_r$, as well as two matrix additions. In contrast, the two-branch convolution performs two straightforward convolutions $W_r \ast X_r$ and $W_i \ast X_i$. Therefore, given the same parameter number, CID-Net provides a better constraint for complex-valued representation than 2BID-Net for more efficient training, and the increased computation requirement would only cause a negligible reduction in inference speed using parallel GPU processing. 

Thirdly, as CID-Net was trained in a supervised manner, the quality of the reconstructed images relied on the quality of the reference images. The results presented show that CID-Net successfully learns better contrast and resolution from the reference obtained by coherently compounding 31 DWs. It can also be seen from Figure \ref{bmode}m that CID-Net eliminates grating lobe artifacts from the input (Fig. \ref{bmode}k). It should nevertheless be noted that standard coherent compounding yields a limited gain in terms of signal-to-noise ratio, as it performs a simple averaging. As standard coherent compounding was used as the reference in the supervised training process, this might hinder further improvement of the proposed method. This can potentially be tackled by including references with better noise quality, such as the line scan quality images used by \cite{zhou2019ultrafast}, or simulated images with ideal image quality. 

Given the reliable effectiveness of the complex-valued network experimentally demonstrated by our results here, we expect that such a tool can similarly benefit I/Q-based US imaging, not only in terms of DW imaging, but also for general US image processing tasks. 
 
\section{Conclusion}

In this paper, a methodology for reconstruction of ultrasound images using CCNN, as CID-Net, is presented. A compounding operator underwent learning to produce high-quality images from I/Q data obtained with a small number of DW transmissions. Experiments were performed on real data from \textit{in-vitro} and \textit{in-vivo} scans. The experimental results show that the proposed CID-Net offers the same image quality as the equivalent real CNN trained with RF data, and outperforms the two-branch CNN architecture that processed the real and imaginary parts of the I/Q signal separately. The proposed work will promote the exploration of CCNN-based approaches for US imaging applications. 

\ifCLASSOPTIONcaptionsoff
  \newpage
\fi

\bibliographystyle{IEEEtran}
\bibliography{IQCompounding.bib}

\end{document}